\documentclass[aip,graphicx]{revtex4-1}
\usepackage{booktabs}
\usepackage{tabularx}
\usepackage{amsmath}  
\usepackage{amssymb}
\usepackage{graphicx} 
\usepackage[margin=1in,letterpaper]{geometry} 
\usepackage[final]{hyperref} 
\usepackage{subcaption}
\usepackage{xcolor} 
\usepackage{float}

\draft 

\begin{document}

\title{Effects of laser-plasma parameters on sub-nanosecond evolution of cross-beam energy transfer}

\author{Yilin Xu}
 \affiliation{School of Science, Shenzhen Campus of Sun Yat-sen University, Shenzhen 518107, China}

\author{Yao Zhao}
 \thanks{Corresponding author}
 \email{zhaoyao5@mail.sysu.edu.cn}
 \affiliation{School of Science, Shenzhen Campus of Sun Yat-sen University, Shenzhen 518107, China}

\author{Hongwei Yin}
 \affiliation{School of Science, Shenzhen Campus of Sun Yat-sen University, Shenzhen 518107, China}
 
\author{Zhuwen Lin}
 \affiliation{School of Science, Shenzhen Campus of Sun Yat-sen University, Shenzhen 518107, China}
 
\author{Yan Yin}
 \affiliation{College of Science, National University of Defense Technology, Changsha 410073, China}
 
\author{Liang Hao}
 \affiliation{Institute of Applied Physics and Computational Mathematics, Beijing 100094, China}

\author{Yaozhi Yi}
 \affiliation{School of Science, Shenzhen Campus of Sun Yat-sen University, Shenzhen 518107, China}

\author{Hongyu Zhou}
 \affiliation{College of Science, National University of Defense Technology, Changsha 410073, China}
 
\author{Jinlong Jiao}
 \thanks{Corresponding author} 
 \email{jiao.jl@zju.edu.cn}
 \affiliation{Zhejiang Institute of Modern Physics, Institute of Astronomy, School of Physics, Zhejiang University, Hangzhou 310027, China}

\author{Anle Lei}
  \affiliation{Shanghai Institute of Laser Plasma, China Academy of Engineering Physics, Shanghai 201800, China}

\begin{abstract}
Cross-beam energy transfer (CBET) between two lasers is investigated through theoretical analysis and two-dimensional hybrid simulations over sub-nanosecond to nanosecond timescales and millimeter spatial scales. A finite frequency-difference range for CBET development is derived. Ion acoustic wave (IAW) damping is found to broaden this range while reducing the growth rate of stimulated Brillouin scattering (SBS). CBET exhibits distinct nonlinear behaviors across different laser-intensity regimes. Denoting $I_{14}=1\times 10^{14}\,\mathrm{W/cm^2}$ for a laser wavelength of $351\,\mathrm{nm}$, at moderate intensities ($1<I/I_{14}<8$), CBET grows weakly and saturates at a low level due to pump depletion. In the strongly coupled SBS regime ($I/I_{14} \gtrsim 8$), harmonic IAW and nonlinear wave-particle interactions emerge. The generation of harmonic IAW reduces the normal IAW mode, while ion-trapping–induced spectral broadening of normal IAW mode causes frequency mismatch, leading to nonlinear reduction of CBET. After the saturation of harmonic mode, ion trapping broadens harmonic IAW spectrum and weakens it, triggering a secondary growth stage of CBET. After approximately $60\,\mathrm{ps}$, CBET approaches quasi-steady-state. The maximum total energy transfer occurs at a frequency difference below the linear matching condition due to the trapping-induced IAW redshift. Based on these two intensity regimes, piecewise scalings of the quasi-saturated total energy transfer rate with $I/I_{14}$ are obtained and shown to be robust against spot size. Speckle effects reduce high-intensity overlap and thus the energy transfer rate. The effects of plasma temperature, density, and flow velocity on CBET are also examined.
\end{abstract}


\maketitle 

\section{Introduction}
After decades of experimental research in inertial confinement fusion (ICF) across both direct-drive and indirect-drive schemes, the control of cross-beam energy transfer (CBET) has been demonstrated as an important issue for achieving high-efficiency ignition\cite{froula2010experimental, pesme2002laser}. CBET is driven by stimulated Brillouin scattering (SBS), which arises when two crossing lasers resonantly excite an ion acoustic wave (IAW) in plasma \cite{macgowan1996laser, michel2009energy}. It leads to laser energy exchange and consequent reductions in implosion velocity and neutron yield\cite{boehly1997initial,campbell1999national,igumenshchev2010crossed,froula2012increasing}. For direct-drive schemes, CBET is treated as the energy loss mechanism \cite{igumenshchev2010crossed}, while it is applied to improve the symmetry of the target implosion \cite{michel2010symmetry,michel2011three,kritcher2021achieving,pickworth2020application,edgell2017mitigation}. In indirect-drive ICF experiments, CBET can impact drive uniformity \cite{michel2009energy, strozzi2017interplay} and reduce the hydrodynamic coupling efficiency \cite{myatt2017wave}. Asymmetrical azimuthal polarization between neighboring beams cause different backscattering in the same cone via CBET \cite{hao2025revealing}. Controlling CBET is an efficient and robust tool to tune the implosion symmetry of ignition capsules\cite{michel2010symmetry,michel2009energy,michel2009tuning}, which is a key to ignition at NIF\cite{kritcher2021achieving,kritcher2022design}.

Previous studies have investigated the effect of several laser parameters on CBET and SBS. In direct-drive conditions, CBET can be mitigated by adjusting the wavelength difference between lasers \cite{glenzer1999thomson}. The intensity of low-frequency beam affects the level and duration of IAW \cite{baldis1996resonant}, while the incident beam intensity ratio determines CBET strength \cite{mckinstrie1996two}. CBET between beams with hot spots in their intensity profiles depends on filament intersections \cite{mckinstrie1998three}. Since CBET is sensitive to beamwidth \cite{mckinstrie1996two}, controlling laser-spot size has been used to improve hydrodynamic efficiency \cite{froula2012increasing}. Multi-color lasers can mitigate CBET by controlling SBS gain \cite{igumenshchev2012crossed}. Broadband lasers with a bandwidth large enough are experimentally proven to be effective in reducing SBS and CBET \cite{lei2024reduction,bates2023suppressing}. However, indirect-drive plasma conditions differ, requiring effective control strategies. It is found that in indirect-drive ICF experiments, the target density in gas-filled hohlraums is on the order of $10^{-2}\,n_c$ ($n_c$ is the critical density where the laser frequency equals the local electron plasma frequency) \cite{gong2019recent,hall2017relationship}. 

Beam smoothing technologies have been developed to control random aberrations and mitigate parametric instabilities in ICF experiments\cite{oudin2025theory}. Among these, continuous phase plates (CPP) \cite{jerome2003design}, polarization smoothing (PS) \cite{boehly1999reduction,Rothenberg2000polarization}, and smoothing by spectral dispersion (SSD) \cite{skupsky1989improved, skupsky1999irradiation, joshua1997comparison} are widely used in modern ICF facilities. It has been proven that a plane wave model overestimates the energy exchange between beams using random phase plates (RPP) crossing at a small angle when only a wavelength shift is present \cite{oudin2021reduction,oudin2022cross,oudin2025theory}. A statistical theory \cite{ruyer2024analytical} shows that forward SBS growth influences CBET by increasing the effective beam aperture \cite{oudin2021reduction}.

Various numerical methods are employed to investigate the mechanism of CBET, including kinetic simulations and fluid simulations \cite{hao2016simulation}. Kinetic simulations, such as Vlasov and PIC methods, capture kinetic and nonlinear effects on picosecond timescales, but are computationally expensive. These methods have been applied to study nonlinear SBS dynamics, saturation mechanisms, and their dependence on laser–plasma conditions \cite{yin2023time,yin2023effects}. Hydrodynamic simulations accurately model regions with strong refraction and are commonly used in implosion studies \cite{zhang2024semi, jia2025modeling}. However, as Hydrodynamic models, they cannot capture kinetic phenomena such as ion trapping and nonlinear frequency shifts \cite{morales1972nonlinear}. For CBET driven by lasers with speckle, hydrodynamic approaches remain computationally efficient. It has been found that ponderomotive self-focusing enhances CBET power transfer for RPP-smoothed beams \cite{huller2020crossed1}, whereas a sufficiently large SSD bandwidth is required to suppress CBET \cite{huller2020crossed2}.

One approach to simulating CBET models electrons as a fluid and ions as particles \cite{vu1996adiabatic}. This method has been validated for accurately and efficiently capturing the nonlinear characteristics of SBS, with or without spatial beam smoothing \cite{cohen1997resonantly,rambo1997hybrid,riconda2000kinetic}. We developed a 2D parallel code, \textbf{HLPI} (\textbf{H}ybrid code of \textbf{L}aser-\textbf{P}lasma \textbf{I}nteractions) \cite{jinlong2022cbetor}, in which the evolution of the laser pulse in plasma is governed by wave equations. The total force on ions is computed from the ion density and laser electric field using Newton’s equation of motion. This force is then applied to update ion positions and velocities. The updated ion positions are then used to compute the new ion density via statistical averaging. This hybrid method preserves ion kinetic accuracy comparable to that of PIC codes, while significantly improving computational efficiency over large spatial and temporal scales.

The saturation mechanisms of CBET and SBS have been extensively studied. Pump depletion governs CBET saturation in a low-ion-heating configuration\cite{hansen2022cross}. Ion trapping and heating modify ion distribution functions, leading to CBET resonance detuning \cite{yin2023time,nguyen2021cross,hansen2022cross}. In addition, the trapping-induced nonlinear frequency shift causes IAW wavefront bowing and breakup, contributing to SBS saturation \cite{williams2004effects,yin2008saturation,yin2023time}. At high density and laser intensity, forward stimulated Raman scattering (FSRS) and backward SBS destabilize the low-frequency beam, resulting in CBET saturation \cite{yin2019saturation, stark2021forward, yin2023time}. Based on these important previous works, we systematically investigate nonlinear effects on CBET evolution in different intensity regimes, particularly ion-trapping-induced IAW spectral broadening and the generation and evolution of harmonic IAW.

In this work, we investigate the linear and nonlinear regimes of CBET both theoretically and numerically. The influence of laser and plasma parameters on CBET is systematically explored. In Section \ref{sec:linear}, we obtain a finite frequency-difference range, within which CBET can be developed. The influence of IAW damping is incorporated. In Section \ref{sec:three}, we first introduce the physical model and boundary conditions in the HLPI code, followed by a benchmarking against a full PIC code. We then perform a series of simulations to investigate the effects of various laser and plasma parameters on CBET, with particular focus on nonlinear ion kinetic effects. We summarize our results in Section \ref{sec:conclusion}.

\section{Theoretical analysis on CBET}
\label{sec:linear}
In this section, we investigate the linear regime of CBET. The theoretical model of multibeam SBS in homogeneous plasmas is studied\cite{zhao2023control}. We consider the dispersion relation of two-color beam interaction with homogeneous plasmas. Let $\omega_{1,2}$ and $\vec{k}_{1,2}$ denote the laser wave frequencies and wave vectors, respectively. The crossing angle of two beams in a vacuum is $\theta$. The fluid equations describing SBS are \cite{kruer2019physics}
\begin{align}
    \left(\frac{\partial^2}{\partial t^2} - c^2 \nabla^2 + \omega_{pe}^2 \right) \vec{\tilde{A}} &= -\frac{4\pi e^2}{m_e} \tilde{n}_e \vec{A}_L \label{eq:fluidEq1}, \\
    \left(\frac{\partial^2}{\partial t^2} - c_s^2 \nabla^2 \right) \tilde{n}_e &= \frac{Zn_0 e^2}{m_e m_i c^2} \nabla^2 \left(\vec{A}_L \cdot \vec{\tilde{A}}\right), \label{eq:fluidEq2}
\end{align}
where $e$ is the charge of electrons, $c$ is the speed of light in a vacuum, $\omega_{pe}$ is the electron plasma frequency, $Z$ is the charge number of ions, and $m_e$ and $m_i$ are the masses of electrons and ions, respectively. $c_s = \sqrt{(Z T_e+3T_i)/m_i}$ is the ion acoustic velocity, $T_e$ and $T_i$ are electron and ion temperatures, respectively. $\vec{A}_L = \vec{A}_{L1} + \vec{A}_{L2}$ is the vector potential of large amplitude light waves, and $\vec{\tilde{A}} =\vec{\tilde{A}}_1 + \vec{\tilde{A}}_2$ and $\tilde{n}_e = n_e - n_0$ are the vector potential of scattered light wave and electron density fluctuation, respectively. 

According to the frequency matching condition, an intense CBET resonance can be found at $\delta\omega_{max} = \omega_2 - \omega_1 = |\vec{k}_2 - \vec{k}_1|c_s = |\delta \vec{k}|c_s \sim 2k_0c_s\sin{\theta/2}$, where $k_0 = (k_1 + k_2)/2$ and $\omega_0=(\omega_1+\omega_2)/2$ is the corresponding central frequency. The relation between normalized intensity $a$ and laser intensity $I$ is $a=\sqrt{I(\mathrm{W/cm^2}) [\lambda(\mathrm{\mu m})]^2/1.37\times10^{18}}$. From Eqs. \eqref{eq:fluidEq1} and \eqref{eq:fluidEq2}, the SBS dispersion relation can be derived \cite{zhao2023control,zhao2021mitigation}. Without loss of generality, we consider small incident angles. The growth rate of SBS is given by the positive imaginary part of the numerical solution of the SBS dispersion relation.

\begin{figure}[t]
    \begin{subfigure}{0.49\textwidth}
        \centering
        \includegraphics[width=\textwidth]{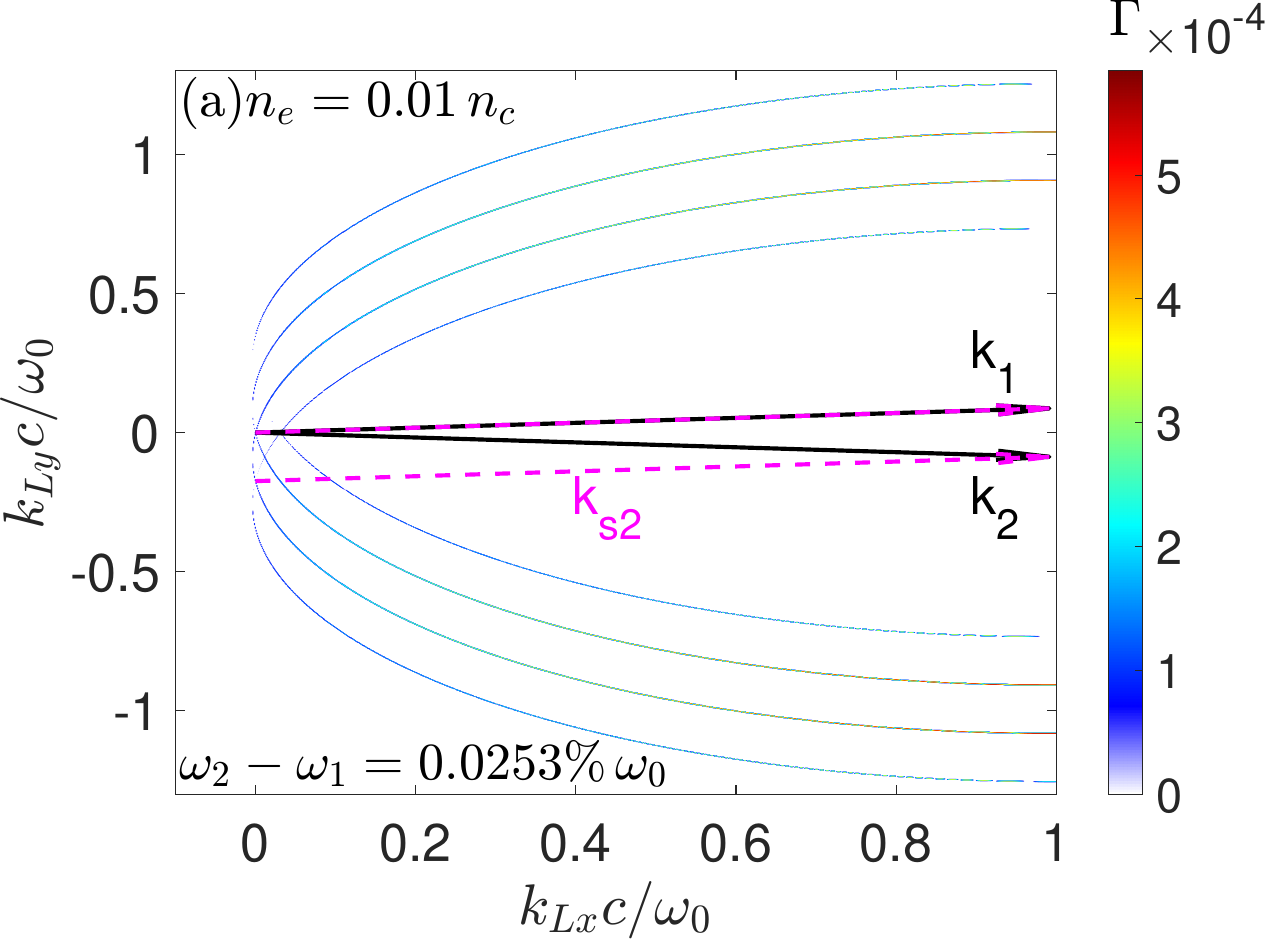}
    \end{subfigure}
    \hfill
    \begin{subfigure}{0.49\textwidth}
        \centering
        \includegraphics[width=\textwidth]{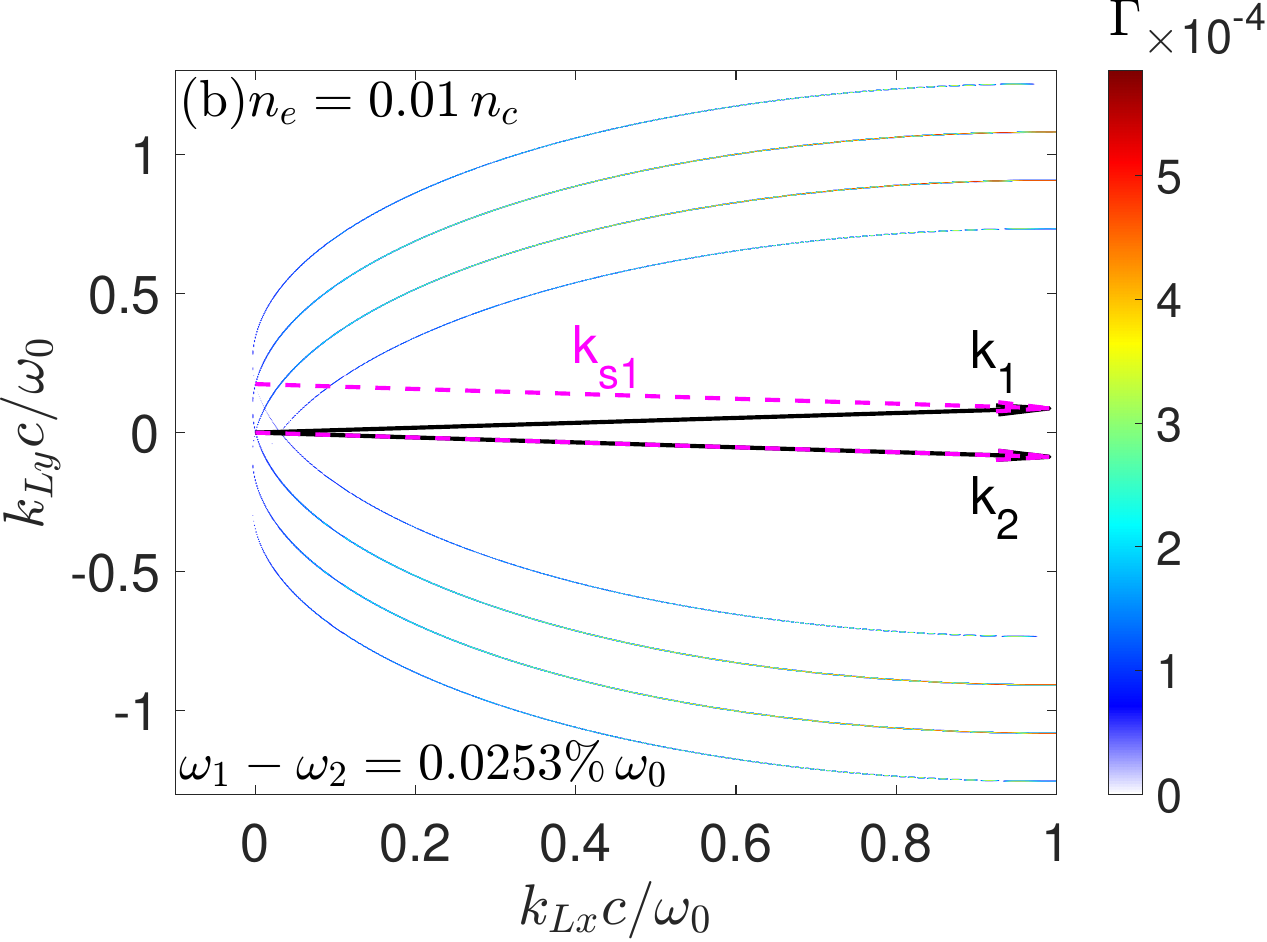}
    \end{subfigure}
    \caption{Numerical solutions of multibeam SBS dispersion relation at different frequency differences (a) $\delta\omega=0.0253\%\,\omega_0$ and (b) $\delta\omega=-0.0253\%\,\omega_0$. $\vec{k}_{sj}$ is the wave vector of the scattering light of the pump beam, where $j=1$ or 2. Normalized intensities are $a_1=a_2=0.027$, crossing angle is $\theta=10^\circ$, electron density is $n_e=0.01\,n_c$, and electron and ion temperatures are $T_e=1\,\mathrm{keV}$ and $T_i=0.333\,\mathrm{keV}$, respectively.}
    \label{fig:k2-k1}
\end{figure}

Figure \ref{fig:k2-k1} shows the process of SBS driven by two laser beams with a frequency difference $\delta\omega=\omega_2-\omega_1$. In addition to the shared IAW, the SBS scattering light driven by the high-frequency beam propagates in the same direction as the low-frequency beam, therefore it is shared by the low-frequency beam. This results in the energy transfer from the high- to the low-frequency beam. We denote the high- and low-frequency beam as pump and seed beam, respectively. In Fig. \ref{fig:k2-k1}(a), the scattering light $\vec{k}_{s2}$ developed by the pump beam $\vec{k}_2$ is shared by the seed beam $\vec{k}_1$, transferring energy from pump beam to seed beam. Conversely, in Fig. \ref{fig:k2-k1}(b), scattering light $\vec{k}_{s1}$ generated by $\vec{k}_1$ is shared by seed beam $\vec{k}_2$, allowing $\vec{k}_2$ to gain energy.

\begin{figure}[t]
    \begin{subfigure}{0.45\textwidth}
        \centering
        \includegraphics[width=\textwidth]{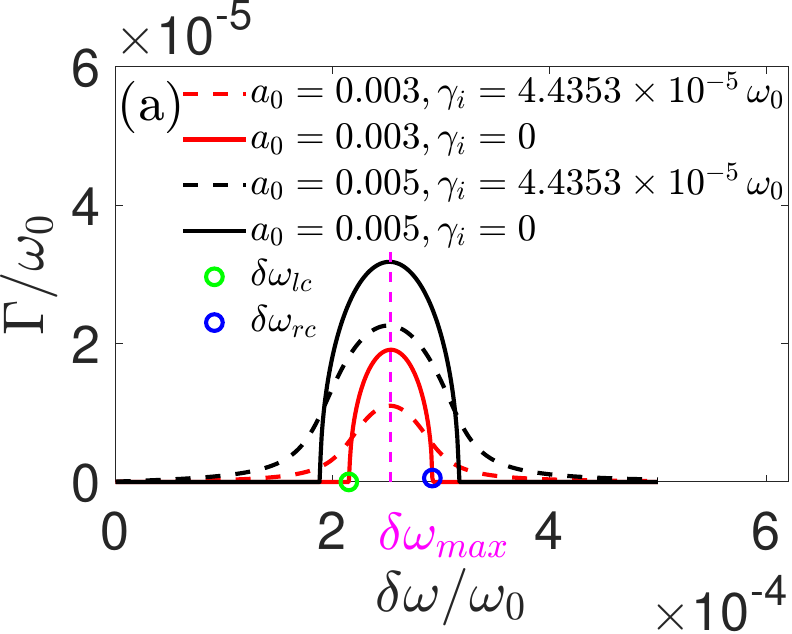}
    \end{subfigure}
    \hfill
    \begin{subfigure}{0.45\textwidth}
        \centering
        \includegraphics[width=\textwidth]{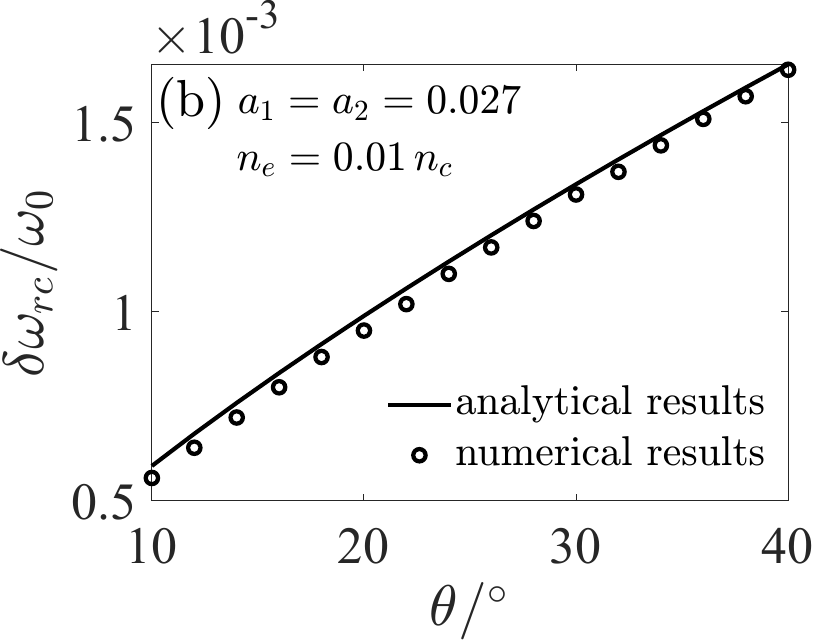}
    \end{subfigure}
    \vspace{0.5cm}
    \begin{subfigure}{0.45\textwidth}
        \centering
        \includegraphics[width=\textwidth]{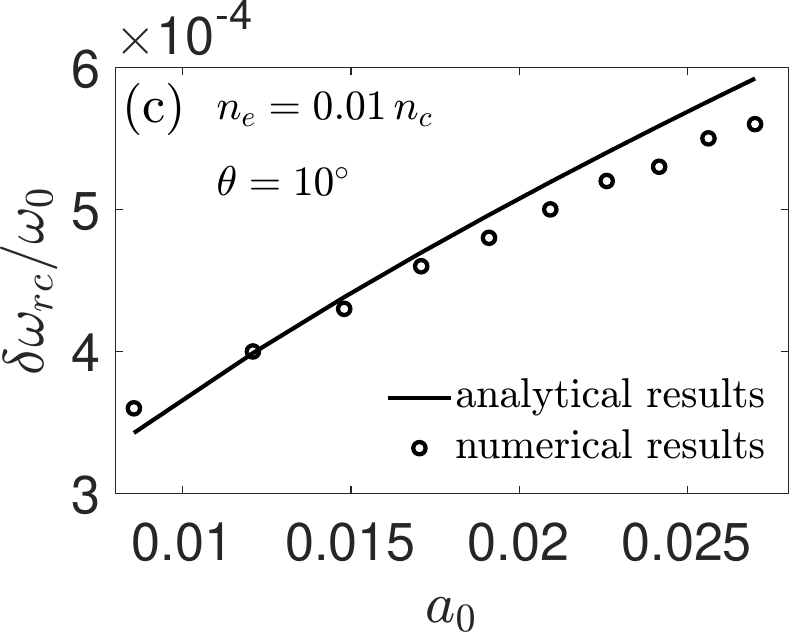}
    \end{subfigure} 
    \hfill
    \begin{subfigure}{0.45\textwidth}
        \centering
        \includegraphics[width=\textwidth]{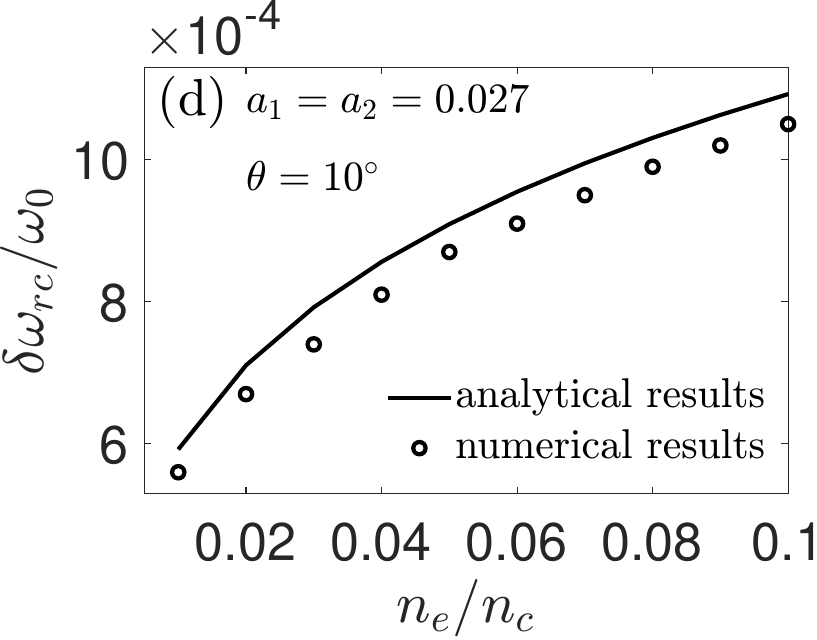}
    \end{subfigure} 
    \caption{(a) The SBS growth rate $\Gamma$ at $\vec{k}_2-\vec{k}_1$ as a function of $\delta\omega$ for $a_0=a_1 = a_2$, with and without damping. Crossing angle is $\theta=10^\circ$ and electron density is $n_e = 0.01\,n_c$. Comparisons of analytical and numerical values of $\delta\omega_{rc}$ as functions of (b) crossing angle $\theta$, (c) normalized intensity $a_0$ and (d) electron density $n_e$. Numerical results are denoted by circles and analytical results are depicted by solid lines. Electron and ion temperatures are $T_e=1\,\mathrm{keV}$ and $T_i=0.333\,\mathrm{keV}$, respectively.}
    \label{fig:cutoff_deltaomega}
\end{figure}

Treating $\vec{k}_2$ as the pump beam, the SBS growth rate $\Gamma$ at $\vec{k}_2 - \vec{k}_1$ is obtained numerically with and without IAW damping. The dependence of $\Gamma$ on the frequency difference $\delta\omega$ is shown in Fig. \ref{fig:cutoff_deltaomega}(a). Plasma conditions are $n_e = 0.01\,n_c, T_e=1\,\mathrm{keV}$ and $T_i=0.333\,\mathrm{keV}$. As indicated, there are left and right cutoff frequency differences $\delta\omega_{lc}$ and $\delta\omega_{rc}$, at which $\Gamma=0$. When $\delta\omega$ falls outside the range $[\delta\omega_{lc}, \delta\omega_{rc}]$, CBET cannot be developed. According to the SBS dispersion relation of two-color beams \cite{zhao2023control}, the theoretical $\delta\omega_{lc}$ and $\delta\omega_{rc}$ for forward SBS (small crossing angle) at low intensities without damping can be obtained as
\begin{align}
    \delta\omega_{lc} &= \sqrt{2}k_0 c_s \sin{\frac{\theta}{2}} - \frac{9a_2^2\omega_{pi}^2c^2}{32c_s^2\omega_1}, \label{left_cutoff_dw}\\
    \delta\omega_{rc} &= k_0c_s\sin{\frac{\theta}{2}}+ \frac{3}{2}\left(\frac{a_1^2\omega_{pi}^2c^2k_0^2\sin^2{\frac{\theta}{2}}}{\omega_1}\right)^{1/3} \label{right_cuttoff_dw}.
\end{align}

The dashed lines in Fig. \ref{fig:cutoff_deltaomega}(a) are obtained by including IAW damping $\gamma_i$ in the SBS dispersion relation \cite{wang2019auto}. The damping modifies the real part of $\omega$ and broadens the frequency-difference range $[\delta\omega_{lc}, \delta\omega_{rc}]$. It also reduces the overall SBS growth rate, thereby introducing a threshold for the development of CBET \cite{zhao2023control}.

Without loss of generality, we consider the effect of right cutoff frequency difference. The analytical and numerical solutions of $\delta\omega_{rc}$ under different conditions of $\theta$, $a_0$, and $n_e$ are shown in Figs. \ref{fig:cutoff_deltaomega}(b)-\ref{fig:cutoff_deltaomega}(d), where the analytical results of Eq. \eqref{right_cuttoff_dw} are in good agreement with the numerical solutions. We find that $\delta\omega_{rc}$ increases with the incident angle, laser intensity and electron density. By tuning these laser-plasma parameters, CBET can be confined within a specific frequency-difference range.

\section{Effects of laser-plasma parameters on CBET}
\label{sec:three}

In this section, the effects of laser-plasma parameters on CBET are investigated using HLPI code. In this code, when lasers propagate in a plasma, the evolution of a laser electric field $\vec{E}$ is described by the wave equation\cite{jinlong2022cbetor}:
\begin{align}
    \label{eq:laser}
    \frac{\partial^2\vec{E}}{\partial t^2} - \nabla^2 \vec{E}+ \nabla(\nabla\cdot\vec{E}) + \omega_{pe}^2\vec{E} = 0.
\end{align}
The low-frequency component of $\vec{E}$ generates the laser ponderomotive force $\vec{F}_p = -1/2 \nabla \langle \vec{E}^2 \rangle$, repelling electrons from ions and forming a charge separation field. When the plasma density has a gradient, the electrons form a thermal pressure, and a self-generated electric field $\vec{E}_s$ appears to balance it:
\begin{align}
    T_e \nabla n_e + e n_e \vec{E}_s = 0,
\end{align}
where electron temperature is assumed constant. Then the ions are pushed by Newton's equation of motion:
\begin{align}
    \frac{\vec{F}_p + Ze\vec{E}_s}{m_i} = \frac{\mathrm{d}\vec{v}}{\mathrm{d}t} = \frac{\mathrm{d}^2\vec{r}}{\mathrm{d}t^2}.
\end{align}
The motion of ions results in change in ion density, leading to the variation of $\omega_{pe}$ and consequently affecting laser field by Eq. \eqref{eq:laser}. The electron density is calculated according to the quasi-neutrality condition ($n_e \approx Zn_i$). 

The boundary for lasers is an open boundary. To avoid the accumulation of hot ions and the loss of a large number of thermal ions observed in conventional ion boundaries, we implement an ion source boundary\cite{jinlong2022cbetor}. The ion source boundary consists of a plasma surrounding the simulation domain and is stationary relative to it, ensuring that no net momentum is introduced. The ion source has the same temperature and density as the initial values of $T_i$ and $n_i$ in the plasma region inside the domain. Ion velocities in this source region are sampled from a two-dimensional Maxwellian distribution corresponding to $T_i$. These ions freely enter the simulation domain, thereby maintaining the desired temperature at the domain edge throughout the simulation. As hot ions leave the domain, the total ion number in simulation is allowed to vary, resulting in a small energy loss at the boundary. However, because a vacuum region separates the boundary from the CBET interaction region, the small variation in the total ion number does not affect CBET behaviors.

\begin{figure}[t]
    \begin{subfigure}{0.47\textwidth}
        \centering
        \includegraphics[width=\textwidth]{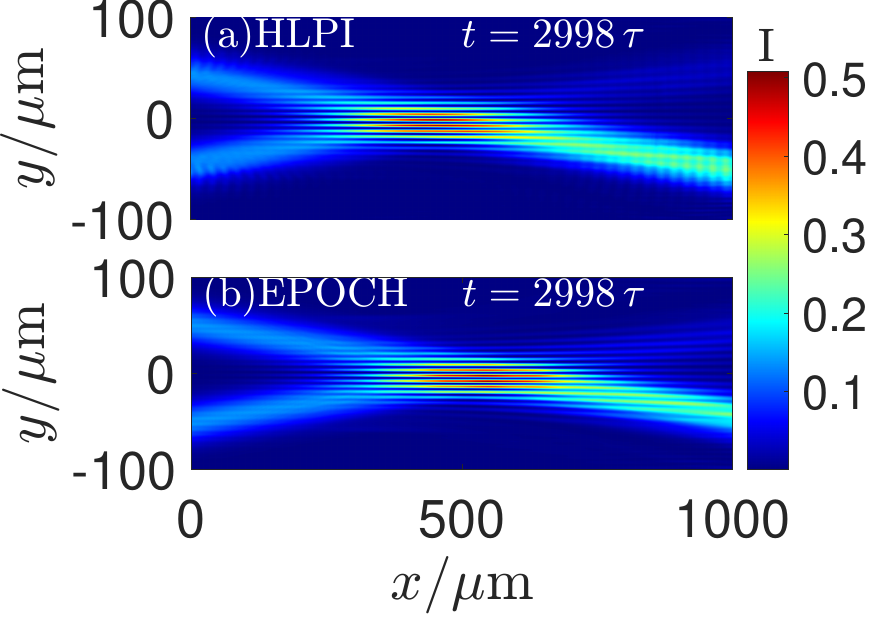}
    \end{subfigure}
    \hfill
    \begin{subfigure}{0.43\textwidth}
        \centering
        \includegraphics[width=\textwidth]{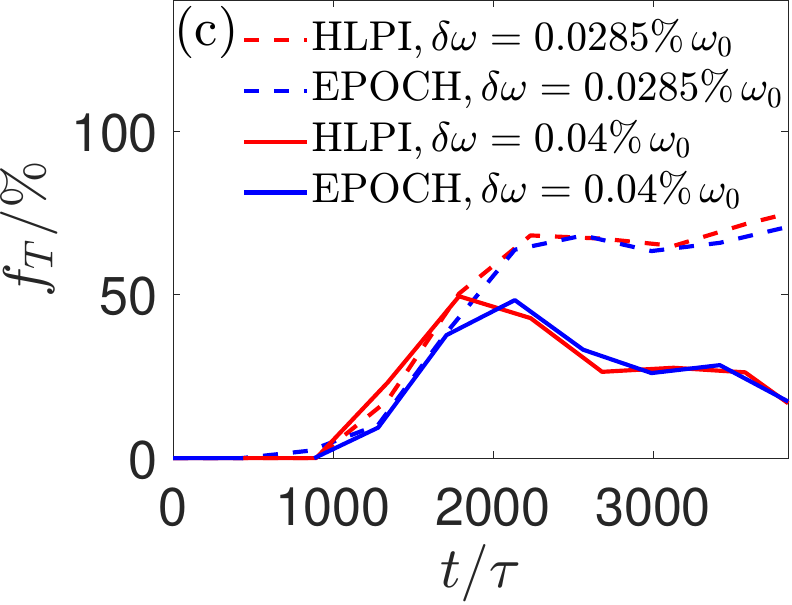}
    \end{subfigure}
    \caption{Benchmarking of the HLPI code against the EPOCH code. Snapshots of laser intensity at $t=2998\tau$ for $\delta\omega=0.026\%\,\omega_0$ obtained from (a) HLPI code and (b) EPOCH code. (c) Comparisons of transient energy transfer rate $f_T$ between two codes for $\delta\omega = 0.0285\%\,\omega_0$ and $0.04\%\,\omega_0$. Main laser-plasma parameters are spot size $w_0=28.96\,\mathrm{\mu m}$, normalized laser intensities $a_1 = a_2 = 0.0232$, crossing angle $\theta=10^\circ$, electron density $n_e = 0.01\,n_c$, electron temperature $T_e = 1\,\mathrm{keV}$ and ion temperature $T_i = 0.333\,\mathrm{keV}$.}
    \label{fig:EPOCHvsHLPI}
\end{figure}

To validate the accuracy of our hybrid code, we benchmark it against full PIC simulations. The spatial resolutions for the PIC code EPOCH are $\Delta x=\Delta y=\lambda/10$, and for the HLPI code are $\Delta x=5\lambda/\pi$ and $\Delta y=\lambda/\pi$. Time step used in the HLPI code is $\Delta t=0.05\,\tau$, where $\tau$ is the laser period corresponding to a laser wavelength of $\lambda=351\,\mathrm{nm}$. The number of particles per cell is $100$. Snapshots of the laser intensity from HLPI and EPOCH under identical laser-plasma conditions are shown in Figs. \ref{fig:EPOCHvsHLPI} (a) and \ref{fig:EPOCHvsHLPI}(b). The corresponding frequency matching condition is $\delta\omega_{max}=0.0253\%\,\omega_0$. As illustrated in Figs. \ref{fig:EPOCHvsHLPI}(a) and \ref{fig:EPOCHvsHLPI}(b), the spatial intensity distributions from the two codes for $\delta\omega=0.026\%\,\omega_0$ at $t=2998\,\tau$ are nearly identical. Therefore, we investigate cases with larger deviations of $\delta\omega$ from $\delta\omega_{max}$ in Fig. \ref{fig:EPOCHvsHLPI}(c).

We define the quasi-steady state and three quantities used to characterize CBET: the transient energy transfer rate $f_T$, the total energy transfer rate $\mathrm{TR}(t)$, and its quasi-saturated value $\mathrm{TR}_s$. At time $t$, we perform a spatial Fourier transform of $E_z$ to separate the two incident lasers in the left and right vacuum regions of the simulation box. The incident and output energies of seed beam are then calculated in the left and right vacuum, respectively. Scattering light is excluded in the calculation of incident energy. The difference between the output and incident energy of seed beam defines the transferred energy. The transient energy transfer rate $f_T$ is the ratio of transferred energy to the incident energy of seed beam at time $t$. The total incident energy of the seed beam is obtained by summing all energy previously injected from the left vacuum, and the total transferred energy is the cumulative transferred energy up to time $t$. Then the total energy transfer rate $\mathrm{TR}(t)$ is defined as the ratio of the total transferred energy to total incident energy of seed beam. We define the quasi-steady state as the regime in which $\mathrm{TR}(t)$ varies by less than $1\%$ over a time interval of $7000\,\tau$. The quasi-saturated value of $\mathrm{TR}(t)$ in this regime is denoted by $\mathrm{TR}_s$. $\mathrm{TR}_s$ depends on simulation duration and is reduced by pump depletion, as well as by smaller spot sizes and transverse domain sizes.

Figure \ref{fig:EPOCHvsHLPI}(c) compares the temporal evolution of $f_T$ for $\delta\omega=0.0285\%\,\omega_0$ and $0.04\%\,\omega_0$ between the EPOCH and HLPI codes, both of which exhibit similar trends. Weak FSRS is observed in EPOCH simulations. The slightly reduced $f_T$ in EPOCH results is attributed to the partial laser energy loss into electron plasma wave (EPW) and scattering light associated with FSRS \cite{stark2021forward}. At such a low electron density, the phase velocity of EPW approaches the speed of light, making efficient electron heating difficult. We do not observe significant electron heating in EPOCH simulations. Langdon effect \cite{turnbull2019langdon} can influence CBET through a shift of the ion-acoustic resonance caused by a non-Maxwellian electron distribution function with exponent $m$. Under the laser-plasma conditions considered in this work, $m$ ranges from $2.04$ to $2.57$. Previous studies show that for $m=3$, the Langdon-induced ion-acoustic frequency upshift is comparable to the ion-trapping-induced down-shift \cite{yin2023time}. Therefore, Langdon effect is expected to play only a minor role in our simulations.

\subsection{Effect of laser parameters on CBET and analysis in nonlinear regime}
\label{sec:nonlinear}
\begin{figure}[htbp]
    \begin{subfigure}{0.42\textwidth}
        \centering
        \includegraphics[width=\textwidth]{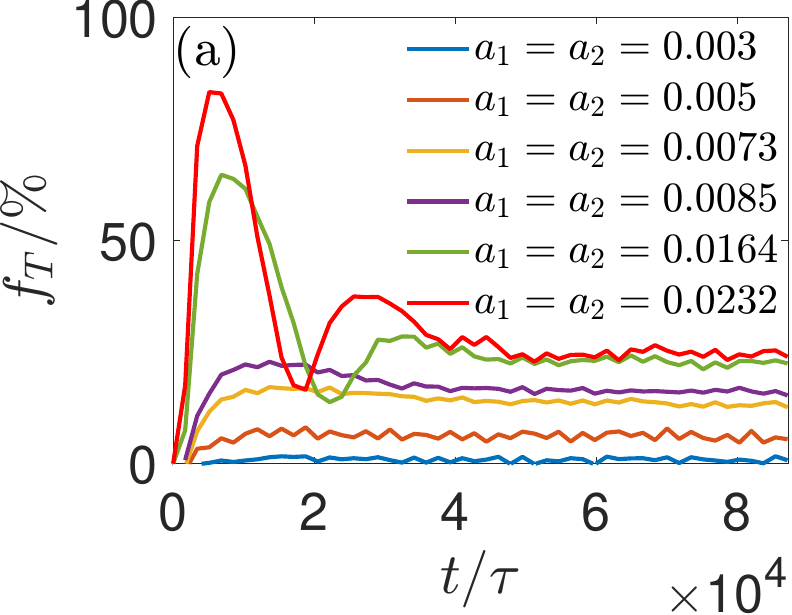}
    \end{subfigure}
    \vspace{0.5cm}
    \begin{subfigure}{0.42\textwidth}
        \centering
        \includegraphics[width=\textwidth]{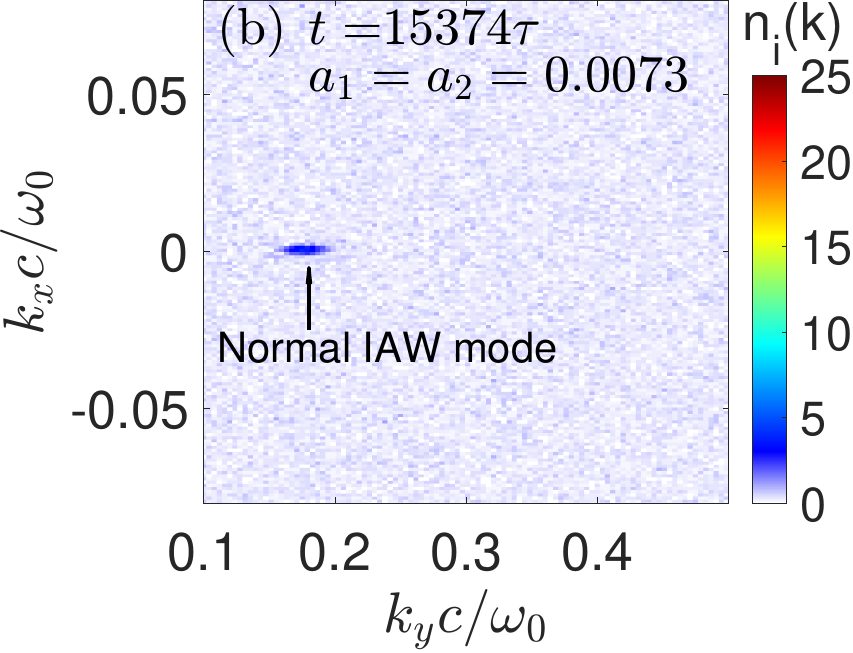}
    \end{subfigure}
    \hfill
    \begin{subfigure}{0.45\textwidth}
        \centering
        \includegraphics[width=\textwidth]{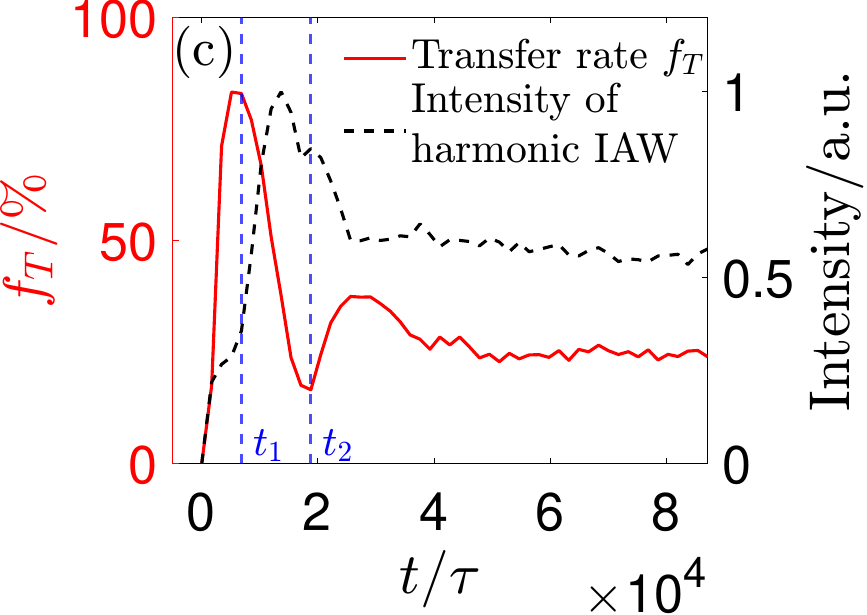}
    \end{subfigure}
    \vspace{0.5cm}
    \begin{subfigure}{0.43\textwidth}
        \centering
        \includegraphics[width=\textwidth]{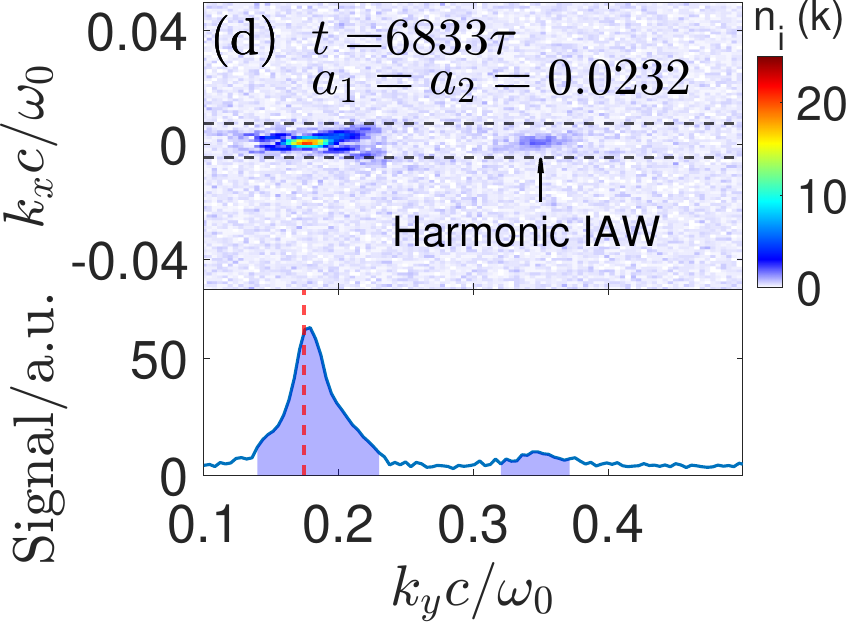}
    \end{subfigure}
    \hfill
    \begin{subfigure}{0.42\textwidth}
        \centering
        \includegraphics[width=\textwidth]{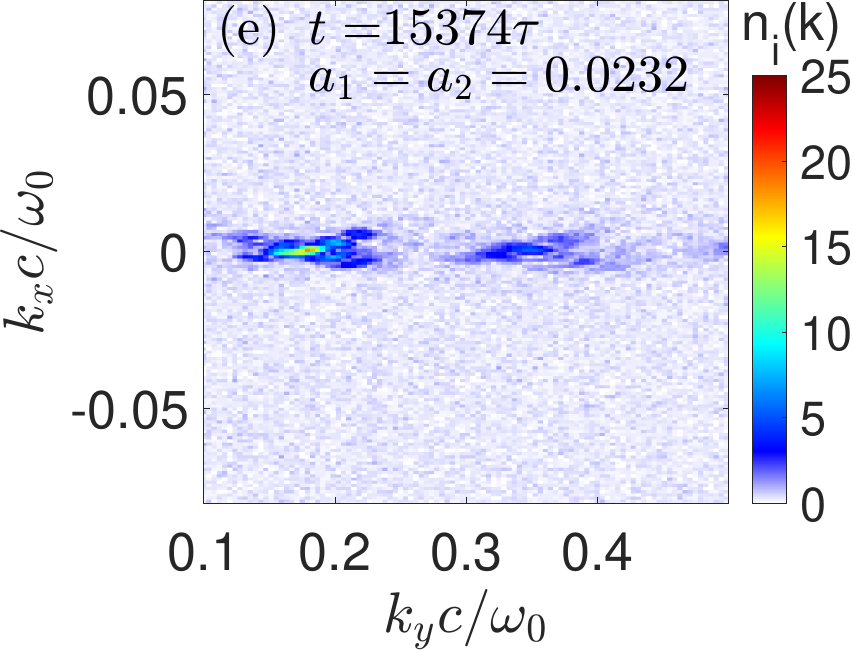}
    \end{subfigure}
    \vspace{0.5cm}
    \begin{subfigure}{0.42\textwidth}
        \centering
        \includegraphics[width=\textwidth]{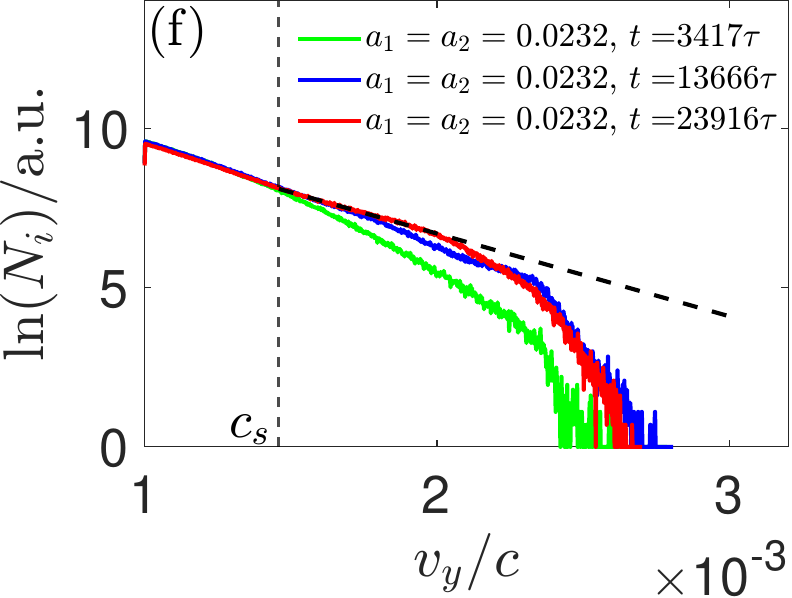}
    \end{subfigure}
    \caption{(a) Temporal evolution of $f_T$ for different laser intensities at $\delta\omega=0.0253\%\,\omega_0$. (b) Ion density distribution in $k$-space at $15374\,\tau$ for $a_1=a_2=0.0073$. For $a_1=a_2=0.0232$: (c) Temporal evolution of $f_T$ and intensity of harmonic IAW. $t_1=6833\,\tau$, $t_2 = 18790\,\tau$. Ion density distributions in $k$-space at $6833\,\tau$ and $15374\,\tau$ are shown in the upper panel of (d), and in (e), respectively. The lower panel of (d) shows the $k_y$-resolved ion density spectral intensity. The red dashed line indicates $k_y c/\omega_0 = 0.17$. (f) Ion $v_y$ distributions for $a_1 = a_2 = 0.0232$ at different times. $N_i$ denotes the relative ion number.}
    \label{fig:deltaomega}
\end{figure}

Several simulations are performed to investigate the effects of laser intensity on CBET. As nonlinear ion kinetic effects have been shown to govern CBET saturation \cite{hansen2022cross, kirkwood2002observation,yin2023effects}, we present a systematic analysis of these effects in this subsection. Plasma parameters are $n_e=0.01\,n_c$, $Z=1$, $T_e=1\,\mathrm{keV}$ and $T_i=0.333\,\mathrm{keV}$. Two lasers with plane wave profiles intersect at an angle of $\theta=10^\circ$ with a pump wavelength of $\lambda_2=351\,\mathrm{nm}$, each with a spot size of $w_0 = 10.17\,\mathrm{\mu m}$. According to the frequency matching condition, we set $\delta\omega = \delta\omega_{max}=0.0253\%\,\omega_0$. The Spatial resolutions and time step are $\Delta x=5\lambda/\pi$, $\Delta y=\lambda/\pi$ and  $\Delta t=0.05\,\tau$, respectively, with $100$ ions per cell. Spatial Fourier transforms are carried out every $2\,\mathrm{ps}$ and the wavenumber-space resolutions are $\Delta k_x = 0.001\, k_0$ and $\Delta k_y = 0.004\, k_0$. The Fourier transform of $n_i$ are carried out over the entire plasma region in the simulation domain.

Figure \ref{fig:deltaomega}(a) displays the temporal evolution of transient energy transfer rate $f_T$ for different values of normalized laser intensities $a_0=a_1=a_2$. At $a_1=a_2=0.003$, CBET is barely observed, as the growth rate at $\delta\omega_{max}$ is too small to drive CBET development due to damping, as shown in Fig. \ref{fig:cutoff_deltaomega}(a). We denote $I_{14}=1\times 10^{14}\,\mathrm{W/cm^2}$ and introduce a normalized laser intensity $I/I_{14}$ for convenience. At moderate intensities ($1<I/I_{14}<8$, corresponding to $0.003<a_0<0.0085$), $f_T$ increases monotonically and saturates at a low level due to pump depletion\cite{hansen2022cross}. Only the amplified normal IAW mode is observed in the wavenumber distribution of ion density at $15374\,\tau$, located at $k_y c/\omega_0 = 0.17$ in Fig. \ref{fig:deltaomega}(b). At this time, CBET has already saturated.

At high intensities ($I/I_{14} \gtrsim 8$, corresponding to $a_0 \gtrsim 0.0085$), CBET enters the strongly coupled SBS regime. Typically, we analyze the case of $a_1=a_2=0.0232$. The SBS growth rate is obtained as $\Gamma=1.4\times10^{-4}\,\omega_0$, corresponding to a characteristic SBS growth time $t_{g} = 1/ \Gamma \approx 1121 \,\tau$ in Fig. \ref{fig:deltaomega}(c). This time is much shorter than the time window for IAW propagating out of the laser crossing region, $t \approx 2w_0/c_s = 39684\,\tau$. The critical plasma length for absolute growth is estimated as $L_{c}=\pi |c_s c_1|^{1/2}/2\Gamma \approx 67\,\lambda_0$, where $c_1$ is the group velocity of SBS scattering light \cite{thomas2011autoresonance}. The transverse size of the laser overlap region is $70\,\lambda_0 > L_c$, and the IAW amplitude is found to grow exponentially, indicating that SBS grows as an absolute instability. We have not found the quasi-mode of IAW in our simulations.

The decrease of $f_T$ from $t_1 = 6833\,\tau$ to $t_2 = 18790\,\tau$ in Fig. \ref{fig:deltaomega}(c) is attributed to the following nonlinear processes. A harmonic IAW mode emerges near $k_y c/\omega_0 = 0.34$ in the upper panel of Fig. \ref{fig:deltaomega}(d) at $t=6833\,\tau$, and grows stronger at $t=15374\,\tau$ in Fig. \ref{fig:deltaomega}(e). In Fig. \ref{fig:deltaomega}(d), the lower-panel signal is obtained by integrating the ion density over the $k_x$ range between the two black dashed lines in the upper $k$-space distribution of $n_i$. The intensities of normal IAW mode and harmonic IAW are obtained by integrating the spectral signal over the shaded regions in the lower panel. As shown in Fig. \ref{fig:deltaomega}(c), harmonic IAW grows from $t_1$, reducing the normal IAW mode. In Fig. \ref{fig:deltaomega}(f), strong ion trapping is observed starting from $t=13666\,\tau$, as the slope of velocity distribution decreases. Ions with velocities around $c_s$ are trapped by IAW and gain energy from it, which leads to the spectral broadening of normal IAW mode in Fig. \ref{fig:deltaomega}(e) and causes a frequency mismatch.

After the harmonic IAW mode saturates, ion trapping broadens its spectrum and reduces it. Ion trapping also nonlinearly modifies the ion distribution function, reducing the IAW damping rate \cite{williams2004effects}. These effects trigger the secondary growth regime of CBET after $t_2=18790\,\tau$ in Fig. \ref{fig:deltaomega}(c). The frequency shift of the IAW induced by ion trapping can lead to bowing and eventual breakup of the wave, leading to the quasi-steady-state saturation of CBET after $50000\,\tau$, corresponding to approximately $60\,\mathrm{ps}$ \cite{yin2008saturation,williams2004effects, yin2019saturation}.

\begin{figure}[t]
    \begin{subfigure}{0.45\textwidth}
        \centering
        \includegraphics[width=\textwidth]{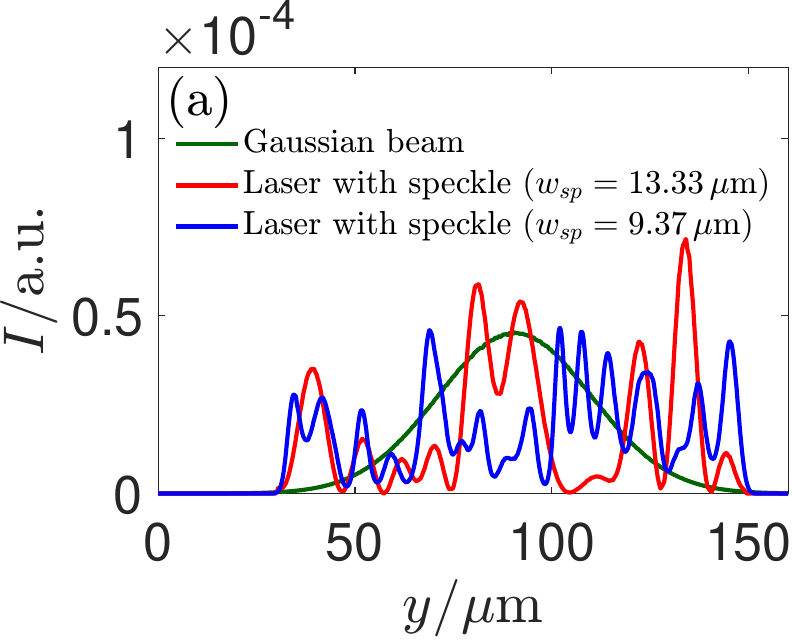}
    \end{subfigure}
    \vspace{0.5cm}
    \begin{subfigure}{0.45\textwidth}
        \centering
        \includegraphics[width=\textwidth]{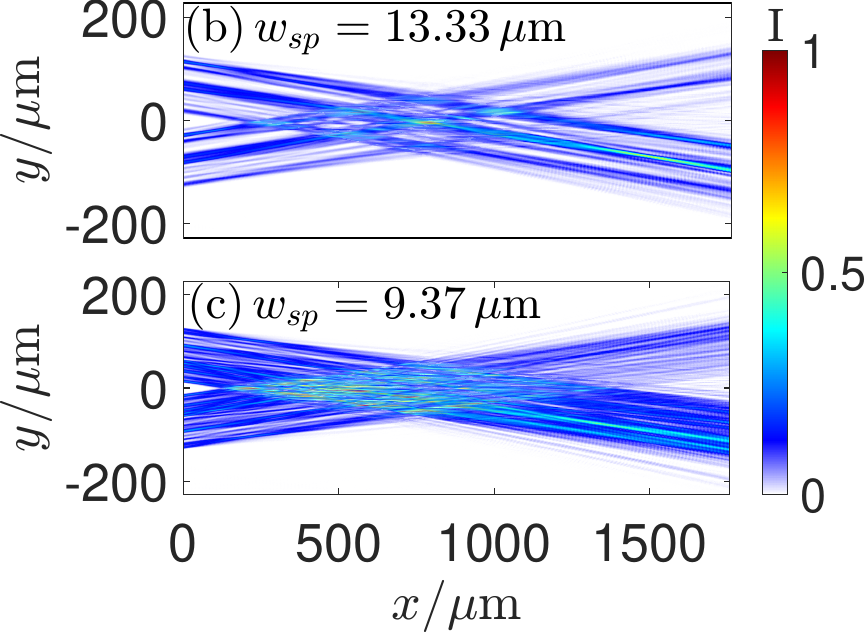}
    \end{subfigure}
    \hfill
    \begin{subfigure}{0.45\textwidth}
        \centering
        \includegraphics[width=\textwidth]{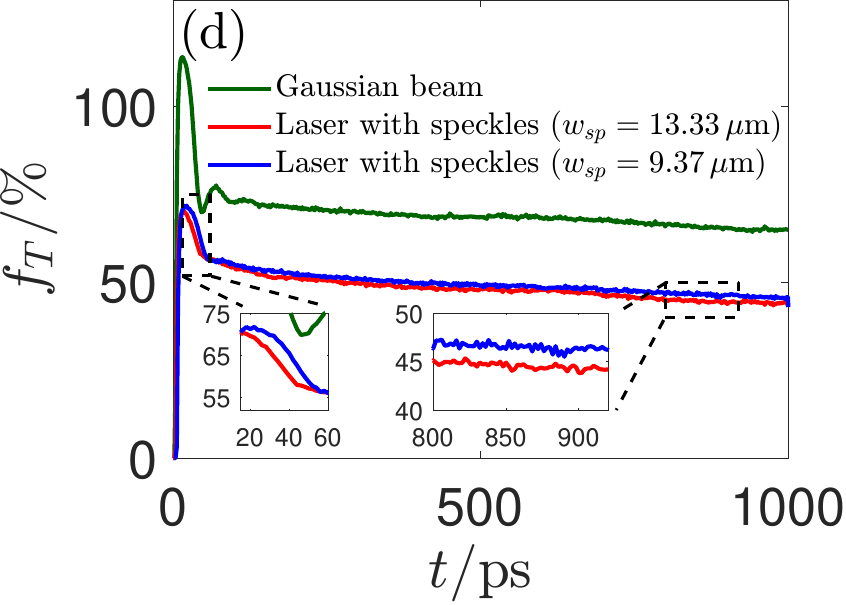}
    \end{subfigure}
    \vspace{0.5cm}
    \begin{subfigure}{0.42\textwidth}
        \centering
        \includegraphics[width=\textwidth]{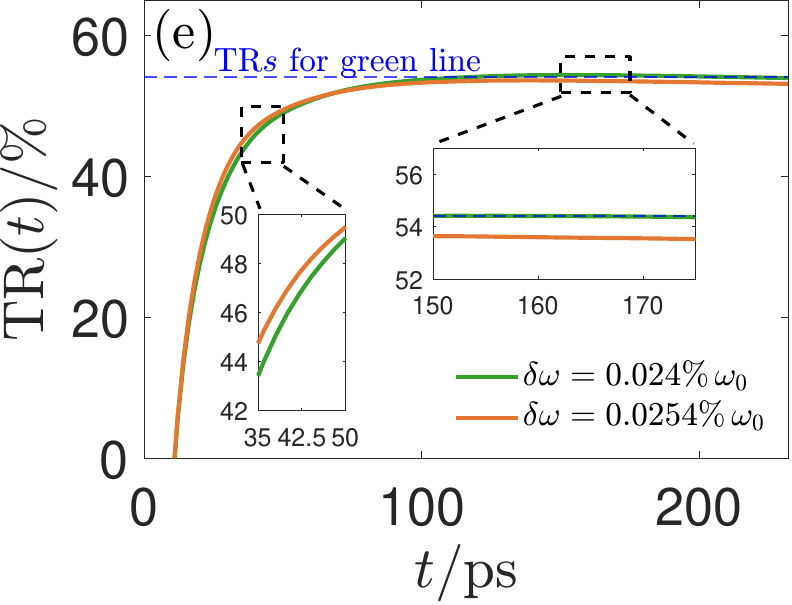}
    \end{subfigure}  
    \caption{(a) Transverse intensity profiles of the Gaussian beam and speckled lasers. Spatial intensity distributions of speckled lasers for speckle widths (b) $w_{sp} = 13.33\,\mathrm{\mu m}$ and (c) $9.37\,\mathrm{\mu m}$ at $t=80\,\mathrm{ps}$. (d) Temporal evolution of $f_T$ for the Gaussian beam and speckled lasers over a duration of $1\,\mathrm{ns}$. (e) Temporal evolution of $\mathrm{TR}(t)$ for speckled lasers with $w_{sp} = 13.33\,\mathrm{\mu m}$ at different $\delta\omega$ over a duration of $232\,\mathrm{ps}$. }
    \label{fig:speckle}
\end{figure}

We then investigate the effect of laser speckle and frequency difference on CBET. The simulation parameters are $\Delta x=\Delta y=\lambda/\pi$ and $\Delta t=0.05\,\tau$, with $50$ ions per cell. The average intensities of two different lasers with speckle widths of $w_{sp} = 13.33\,\mathrm{\mu m}$ and $9.37\,\mathrm{\mu m}$ are $4\times 10^{14}\,\mathrm{W/cm^2}$. The energy of speckled lasers is equal to that of a Gaussian beam with intensity $1\times10^{15}\,\mathrm{W/cm^2}$ and spot size $w_0=45\,\mathrm{\mu m}$. Figure \ref{fig:speckle}(a) shows the transverse intensity distributions of the speckled lasers and the corresponding Gaussian beam. Figures \ref{fig:speckle}(b) and \ref{fig:speckle}(c) show the spatial intensity distributions at $80\,\mathrm{ps}$ for speckled lasers. Plasma parameters are $0.01\,n_c$, $T_e = 1\,\mathrm{keV}$ and $T_i=0.333\,\mathrm{keV}$. Crossing angle is $10^\circ$ and frequency difference is $0.026\%\,\omega_0$, which is close to $\delta\omega_{max} = 0.0253\%\,\omega_0$.

The comparison of transient energy transfer rate $f_T$ between the Gaussian beam and speckled lasers with different $w_{sp}$ is shown in Fig. \ref{fig:speckle}(d). Compared with the Gaussian beam, CBET between two speckled lasers is considerably weaker. This can be attributed to the reduced effective overlap region of high intensities for speckled lasers compared with Gaussian beam\cite{oudin2021reduction}. After the decline, $f_T$ of speckled lasers reaches a quasi-steady state. Compared with lasers with $w_{sp} = 13.33\,\mathrm{\mu m}$, the transient energy transfer rate $f_T$ is slightly higher for lasers with $w_{sp} = 9.37\,\mathrm{\mu m}$, which can be attributed to the enlarged effective overlap region. We adopt the threshold intensity $I_{th}$, defined by $I_{th} \lambda_0^2 \equiv 10^{14}\,\mathrm{W/cm^2 \,\mu m^2}$. For $\lambda_0=351\,\mathrm{nm}$, this corresponds to $I_{th} = 8.16\times 10^{14}\,\mathrm{W/cm^2}$. When average intensity $I_{ave} < 0.75 I_{th} \approx 6 \times 10^{14}\,\mathrm{W/cm^2}$, speckle plays a role only in their abundance in the crossing region \cite{colaitis2016modeling, huller2017impact}. This average intensity is also below the threshold $1.62\times 10^{15}\,\mathrm{W/cm^2}$ for the onset of ponderomotive self-focusing at $\lambda_0=351\,\mathrm{nm}$ and $T_e = 1\,\mathrm{keV}$ \cite{huller2020crossed1}. Although the speckle exhibit high intensities, the nonlinear effects they induce lead to a quasi-saturated CBET gain, so that $f_T$ does not increase significantly.

\begin{table}[t]
    % \begin{ruledtabular}
    \caption{\label{tab:total_fT_speckle} $\mathrm{TR}_s$ for lasers with $w_{sp}=13.33\,\mathrm{\mu m}$ and $I_{ave} = 4\times10^{14}\,\mathrm{W/cm^2}$ at different $\delta\omega$, up to $t=236\,\mathrm{ps}$.}
    \begin{tabular}{c@{\hspace{3cm}}c}
        \hline\hline
        $\delta\omega$ & $\mathrm{TR}_s$\\
        \hline
        $0.01\%\,\omega_0$ & $8.2496\%$\\
        \hline
        $0.020\%\,\omega_0$ & $43.9993\%$\\
        \hline
        $0.022\%\,\omega_0$ & $52.3415\%$\\
        \hline
        $0.024\%\,\omega_0$ & $56.6443\%$\\
        \hline
        $0.0254\%\,\omega_0$ & $54.6753\%$\\
        \hline
        $0.026\%\,\omega_0$ & $54.4920\%$\\
        \hline
        $0.028\%\,\omega_0$ & $47.0484\%$\\
        \hline
        $0.030\%\,\omega_0$ & $37.1265\%$\\
        \hline
        $0.04\%\,\omega_0$ & $4.5624\%$\\
        \hline
        $0.05\%\,\omega_0$ & $1.1630\%$\\
        \hline\hline
    \end{tabular}
    % \end{ruledtabular}
\end{table}

The evolution of $\mathrm{TR}(t)$ for $\delta\omega=0.024\%\,\omega_0$ and $0.0254\%\,\omega_0$ with $w_{sp}=13.33\,\mathrm{\mu m}$ is shown in Fig. \ref{fig:speckle}(e). Over a duration of $232\,\mathrm{ps}$, $\mathrm{TR}(t)$ increases and reaches the quasi-steady state. In the linear regime, $\mathrm{TR}(t)$ of $\delta\omega = 0.0254\%\,\omega_0$ is higher than that of $0.024\%\,\omega_0$, since $0.0254\%\,\omega_0$ is closer to $\delta\omega_{max}$. However, in the nonlinear regime, due to the redshift of IAW frequency induced by ion trapping, $\mathrm{TR}_s$ for $\delta\omega=0.0254\%\,\omega_0$ becomes lower than that of $0.024\%\,\omega_0$. Table \ref{tab:total_fT_speckle} lists $\mathrm{TR}_s$ obtained over a duration of $236\,\mathrm{ps}$ for different frequency differences. Under the same conditions, the theoretical left and right cutoff frequency differences for the Gaussian beam case in Fig. \ref{fig:speckle}(a) are obtained as $\delta\omega_{lc} = 0.012\%\,\omega_0$ and $\delta\omega_{rc} = 0.037\%\,\omega_0$. Accordingly, $\mathrm{TR}_s$ for speckled lasers of $\delta\omega=0.01\%\,\omega_0$ and $\delta\omega=0.04\%\,\omega_0$ correspond to only $14.6\%$ and $8\%$ of the maximum $\mathrm{TR}_s$ value in Table \ref{tab:total_fT_speckle}, respectively. The $\mathrm{TR}_s$ for $\delta\omega$ ranging from $0.020\%\,\omega_0$ to $0.030\%\,\omega_0$ can be fitted by a quadratic function.

\begin{figure}[t]
    \centering
    \includegraphics[width=0.5\textwidth]{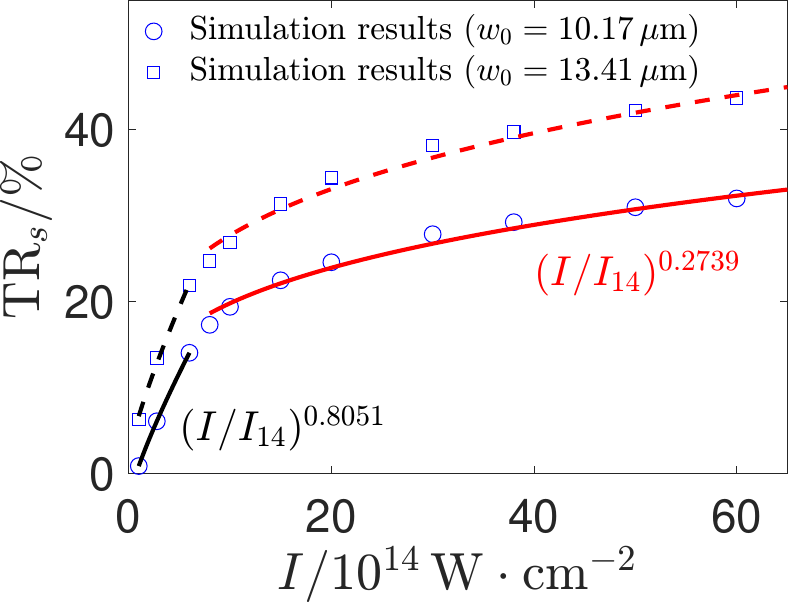}
    \caption{Scalings of $\mathrm{TR}_s$ with plane-wave normalized laser intensity $I/I_{14}$. Blue circles and squares denote simulation results for laser spot sizes of $w_0=10.17\,\mathrm{\mu m}$ and $13.41\,\mathrm{\mu m}$, respectively. Solid and dashed lines represent the corresponding fitting curves for the two spot sizes. Black lines follow a scaling of $(I/I_{14})^{0.8051}$, while red lines follow a scaling of $(I/I_{14})^{0.2739}$. Plasma conditions are $n_e=0.01\,n_c$, $T_e=1\,\mathrm{keV}$ and $T_i=0.333\,\mathrm{keV}$.}
    \label{fig:scaling_I}
\end{figure}

Finally, we present the scalings of $\mathrm{TR}_s$ with the normalized laser intensity $I/I_{14}$ for plane wave laser profiles in Fig. \ref{fig:scaling_I}. As the dominant nonlinear effects differ between moderate intensity regime ($1<I/I_{14}<8$) and the strongly coupled SBS regime ($I/I_{14} \gtrsim 8$), piecewise scaling laws of $\mathrm{TR}_s$ are obtained as a function of $I/I_{14}$. The scaling exponents remain consistent when the laser spot size is increased to $w_0=13.41\,\mathrm{\mu m}$, indicating that the scaling law is robust against the laser spot size.

\subsection{Effects of plasma parameters on CBET}
\label{sec:plasma_parameter}
\begin{figure}[htbp]
    \begin{subfigure}{0.4\textwidth}
        \centering
        \includegraphics[width=\textwidth]{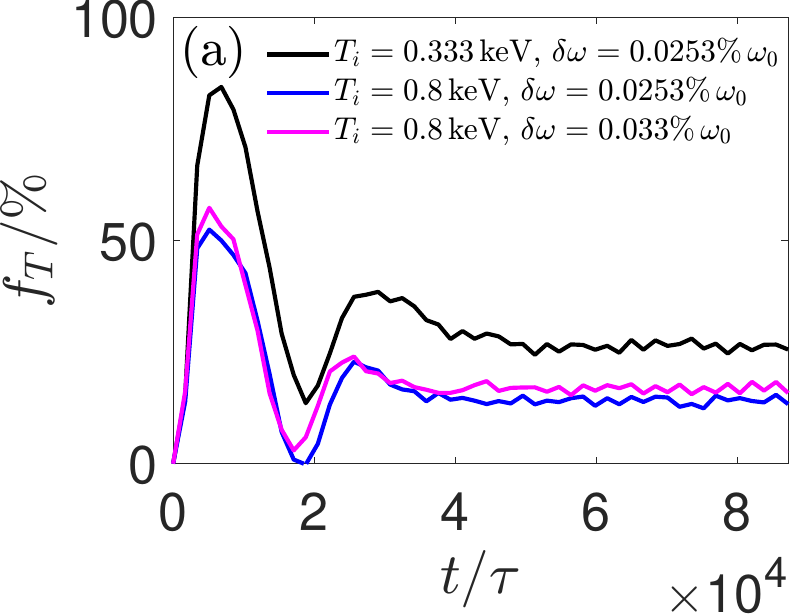}
    \end{subfigure}
    \hfill
    \begin{subfigure}{0.4\textwidth}
        \centering
        \includegraphics[width=\textwidth]{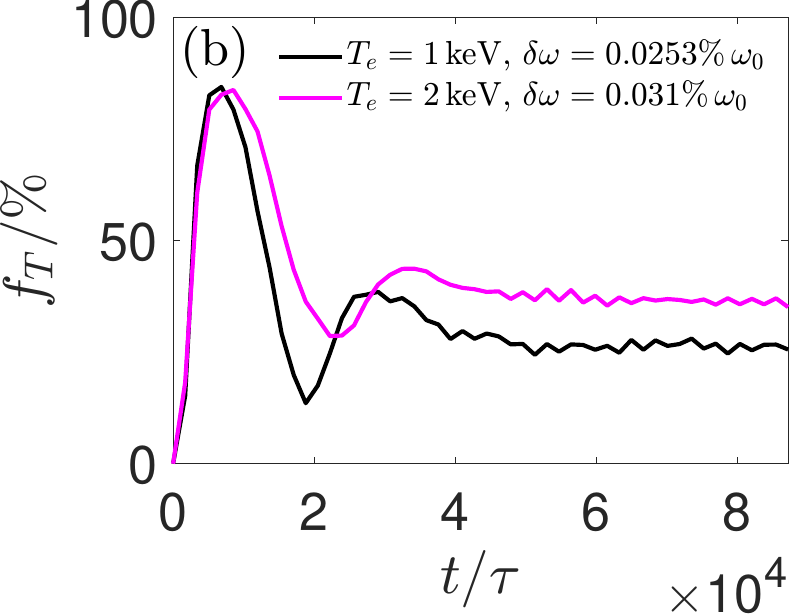}
    \end{subfigure}
    \vspace{0.5cm}
    \begin{subfigure}{0.4\textwidth}
        \centering
        \includegraphics[width=\textwidth]{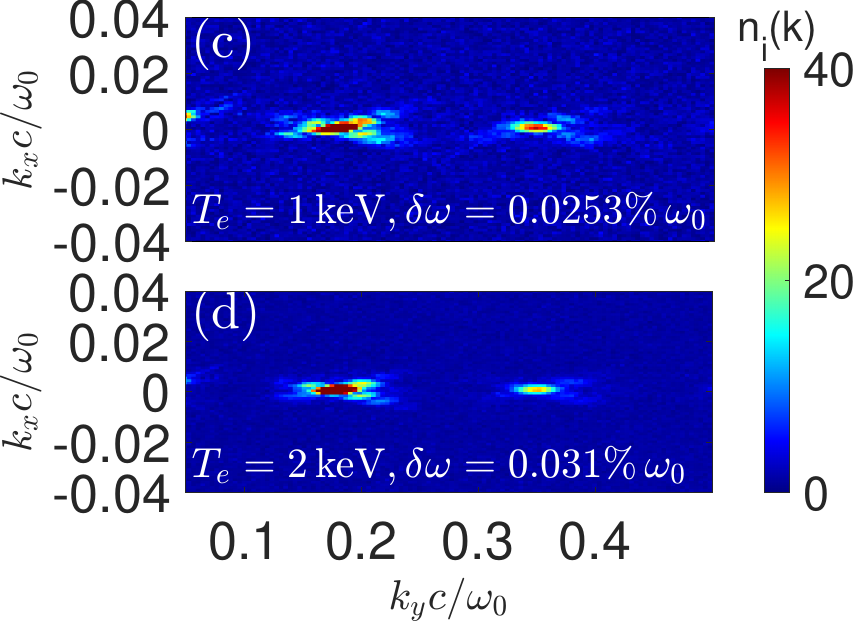}
    \end{subfigure}
    \hfill
    \begin{subfigure}{0.4\textwidth}
        \centering
        \includegraphics[width=\textwidth]{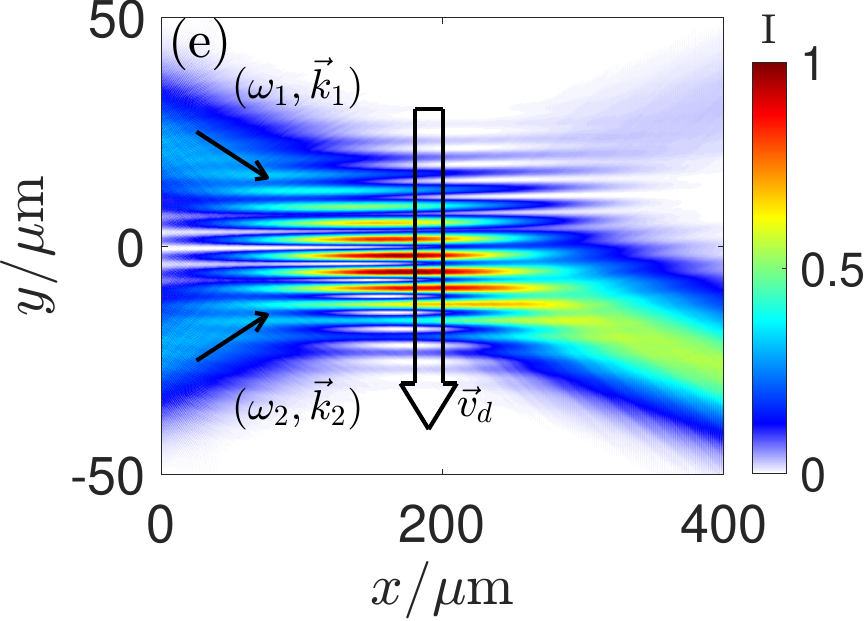}
    \end{subfigure}
    \vspace{0.5cm}
    \begin{subfigure}{0.4\textwidth}
        \centering
        \includegraphics[width=\textwidth]{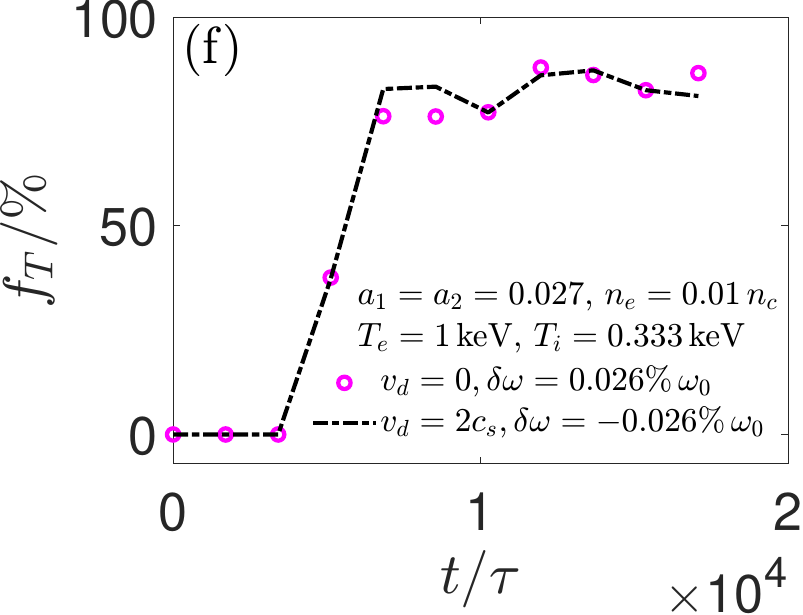}
    \end{subfigure}
    \hfill
    \begin{subfigure}{0.4\textwidth}
        \centering
        \includegraphics[width=\textwidth]{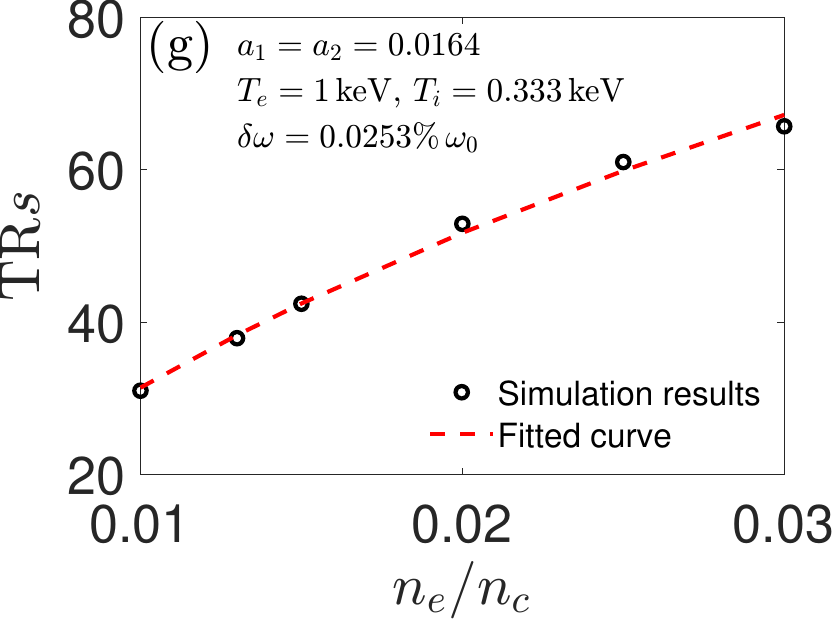}
    \end{subfigure}
    \caption{(a) Temporal evolution of $f_T$ for different $T_i$ with $T_e = 1\,\mathrm{keV}$. $\delta\omega_{max}=0.033\%\,\omega_0$ corresponds to $T_i=0.8\,\mathrm{keV}$. (b)Temporal evolution of $f_T$ for different $T_e$ with $T_i=0.333\,\mathrm{keV}$. $\delta\omega_{max}=0.031\%\,\omega_0$ corresponds to $T_e=2\,\mathrm{keV}$. Normalized intensities are $a_1=a_2=0.0212$ and electron density is $n_e = 0.01\,n_c$. Ion density distributions in $k$-space over $[6833, 13666]\tau$ for (c) $T_e=1\,\mathrm{keV}$, $\delta\omega=0.0253\%\,\omega_0$ and (d) $T_e=2\,\mathrm{keV}$, $\delta\omega=0.031\%\,\omega_0$. (e) Spatial distribution of laser intensity with a plasma flow $v_d = 2c_s$ and frequency difference $\delta\omega = -0.026\%\,\omega_0$. (f) Temporal evolution of $f_T$ for the cases with $\delta\omega=0.026\%\,\omega_0$ only and with both $\delta\omega=-0.026\%\,\omega_0$ and $v_d = 2 c_s$. (g) Scaling of $\mathrm{TR}_s$ with respect to $n_e$.}
    \label{fig:Tem}
\end{figure}

In this subsection, we investigate the effect of plasma parameters on CBET, including initial electron temperature $T_e$, ion temperature $T_i$, transverse flow velocity $\vec{v}_d$ and electron density $n_e$. All simulations in this subsection are based on plane wave lasers with the pump beam wavelength $\lambda_2=351\,\mathrm{nm}$ and a crossing angle of $\theta=10^\circ$. The simulation parameters are $\Delta x=5\lambda/\pi$, $\Delta y=\lambda/\pi$, and  $\Delta t=0.05\,\tau$, with $100$ ions per cell. 

Figures \ref{fig:Tem}(a) and \ref{fig:Tem}(b) clearly demonstrate how the transient energy transfer rate varies with $T_i$ and $T_e$, respectively. The normalized laser intensity is $a_1 = a_2 = 0.0212$ and electron dnesity is $n_e = 0.01\,n_c$. All results are evaluated relative to the reference case with $T_e=1\,\mathrm{keV}$, $T_i=0.333\,\mathrm{keV}$ and $\delta\omega=0.0253\%\,\omega_0$. Changes in plasma temperature alter the ion-acoustic velocity, thereby modifying the frequency matching condition $\delta\omega_{max}$ and affecting CBET.

As shown in Fig. \ref{fig:Tem}(a), increasing $T_i$ from $0.333\,\mathrm{keV}$ to  $0.8\,\mathrm{keV}$ decreases both the linear saturation level of $f_T$ and its quasi-steady-state value. This trend agrees with the theoretical prediction that the SBS growth rate in the reference case is higher than that for $T_i = 0.8\,\mathrm{keV}$ at the same $\delta\omega$. A lower $T_e/T_i$ leads to stronger IAW damping, which accounts for the decrease of the quasi-steady-state value of $f_T$ as $T_i$ increases \cite{yin2023time}. When frequency matching condition is satisfied, the linear saturation and quasi-steady-state value of magenta curve are slightly higher than those of blue curve.

Figure \ref{fig:Tem}(b) shows the influence of $T_e$. For $\delta\omega = \delta\omega_{max}$, two $f_T$ curves nearly overlap in the linear growth regime, indicating that $T_e$ influences CBET mainly through the frequency matching condition at this stage. Increasing $T_e$ yields a larger quasi-steady-state $f_T$ value, which can be attributed to reduced damping associated with the increased $T_e/T_i$ ratio. From $t = 6833\,\tau$ to $13666\,\tau$, the harmonic IAW is weaker at $T_e = 2\,\mathrm{keV}$ than at $T_e = 1\,\mathrm{keV}$, as shown in Figs. \ref{fig:Tem}(d) and \ref{fig:Tem}(c), respectively. As discussed in section \ref{sec:nonlinear}, the weaker reduction of $f_T$ in the nonlinear regime results from the weaker harmonic IAW.

The transverse flow velocity $\vec{v}_d$ affects CBET through the linear matching condition $\omega_2 - \omega_1 = \delta\vec{k} \cdot \vec{v}_d + |\delta\vec{k}|c_s$. In the case of a transverse sonic plasma flow, the Doppler shift introduces a local frequency difference in the plasma rest frame \cite{myatt2017wave}. Here, $\vec{v}_d$ is directed along the negative $y$-axis in Fig. \ref{fig:Tem}(e). Figure \ref{fig:Tem}(f) compares the effects of frequency difference and flow velocity on CBET. Circles represent cases with only $\delta\omega = 0.026\%\,\omega_0$, while the black dashed line corresponds to the case with $\delta\omega=-0.026\%\,\omega_0$ and $v_d = 2c_s$. In both cases, the linear matching condition is satisfied, and the beam with $(\omega_1, \vec{k}_1)$ gains energy, even when it is the high-frequency beam. This result indicates that when $v_d$ is sufficiently large and the matching condition is met, energy can be transferred from the low-frequency beam to the high-frequency beam.

The scaling of $\mathrm{TR}_s$ with the initial electron density $n_e$ is presented in Fig. \ref{fig:Tem}(g). Based on the SBS dispersion relation of two-color beams \cite{zhao2023control}, the SBS growth rate scales as $\Gamma \propto \sqrt{n_e}(1 - n_e)^{1/4}$. This scaling agrees well with the simulated values of $\mathrm{TR}_s$.

\section{Conclusion}
\label{sec:conclusion}
In this work, we investigate the linear and nonlinear evolution of CBET under indirect-drive conditions over sub-nanosecond to nanosecond timescales and millimeter spatial scales. Analysis of multi-beam SBS indicates a finite frequency-difference range for the development of CBET. IAW damping broadens this range and reduces SBS growth rate, introducing a threshold for CBET development, which is examined numerically.

We employ a hybrid code HLPI with mobile ions and Boltzmann electrons, which is accurate and more efficient compared with the full PIC code. Laser intensity $I$ plays a critical role in determining the behavior of CBET, influencing the dominant nonlinear effect on CBET saturation. We define a normalized laser intensity $I/I_{14}$, where $I_{14} = 1\times 10^{14}\,\mathrm{W/cm^2}$. At moderate normalized laser intensity regime ($1<I/I_{14}<8$), pump depletion causes CBET saturates at a low level after the linear growth. In the strongly coupled SBS regime ($I/I_{14} \gtrsim 8$), harmonic IAW and ion trapping dominate the nonlinear evolution of CBET. The generation and growth of harmonic IAW reduces the energy of normal IAW mode. Ion trapping induces the spectral broadening of normal IAW mode, causing frequency mismatch. These effects lead to the nonlinear reduction of CBET. After the saturation of harmonic IAW mode, ion trapping reduces damping rate, broadens harmonic IAW spectrum and weakens it, resulting in the secondary growth stage of CBET. CBET approaches quasi-steady-state saturation due to IAW bowing and breakup after approximately $60\,\mathrm{ps}$. The piecewise scaling exponents of the quasi-saturated total energy transfer rate with respect to $I/I_{14}$ are $0.8051$ at moderate intensities and $0.2739$ in strongly coupled SBS regime, which are robust against laser spot size.

Using CPP smoothed lasers from indirect-drive ICF experiments, we investigate the influence of speckle structure on CBET. Compared with Gaussian beams, the reduced effective overlap region of high intensities results in a lower transient energy transfer rate for speckled lasers. A lower speckle width can enhance CBET slightly. The frequency-difference range for CBET development is examined using speckled lasers, which agrees well with theoretical expectations. The maximum total energy transfer occurs at a frequency difference below the frequency matching condition $\delta\omega_{max}$, which is attributed to nonlinear redshift of IAW frequency induced by ion trapping. 

A higher ion temperature leads to stronger IAW damping, which mitigates CBET in both linear and nonlinear regimes, regardless of whether the frequency matching condition is satisfied. The electron temperature influences CBET mainly through the frequency matching condition in the linear regime. However, in the nonlinear regime, an increase in electron temperature weakens harmonic IAW, leading to a higher quasi-steady-state level. The quasi-saturated total energy transfer rate increases with electron density $n_e$, following a scaling of $\sqrt{n_e}(1-n_e)^{1/4}$. A transverse plasma drift mainly influence the frequency matching condition in the linear regime. When the drift velocity is sufficiently large, it enables energy transfer from the low-frequency beam to the high-frequency one.

\begin{acknowledgments}
The authors acknowledge valuable discussion with Lifeng Wang and Hongbo Cai. This work is supported by the National Natural Science Foundation of China (No.12005287). The EPOCH code used in this work was in part funded by the UK EPSRC grants EP/G054950/1, EP/G056803/1, EP/G055165/1, EP/ M022463/1 and EP/P02212X/1.
\end{acknowledgments}

\section*{Author Declarations}
\subsection*{Conflict of interest}
The authors have no conflicts to disclose.

\section*{Data Availability Statement}
The data that supports the findings of this study are available from the corresponding author upon reasonable request.

\bibliography{main}

\end{document}